\documentstyle[epsfig]{mn}

\newcommand{\mnras}{MNRAS}
\newcommand{\apj}{ApJ}

\title[]
{The age dependence of halo clustering}

\author[L. Gao, V. Springel, \& S.~D.~M. White]  {Liang ~Gao\thanks{Email:
        gaoliang@mpa-garching.mpg.de}, Volker Springel, Simon D.~M.~White \\
        Max--Planck--Institut f\"ur Astrophysik, D-85748 Garching,
        Germany}

\begin{document}
\label{firstpage} \maketitle

\begin{abstract}
We use a very large simulation of the concordance $\Lambda$CDM
cosmogony to study the clustering of dark matter haloes. For
haloes less massive than about $10^{13}h^{-1}{\rm M_\odot}$ the
amplitude of the two-point correlation function on large scales
depends strongly on halo formation time. Haloes that assembled at high
redshift are substantially more clustered than those that assembled
more recently. The effect is a smooth function of halo formation time
and its amplitude increases with decreasing halo mass. At
$10^{11}h^{-1}{\rm M_\odot}$ the ``oldest'' 10\% of haloes are more
than 5 times more strongly correlated than the ``youngest'' 10\%. This
unexpected result is incompatible with the standard excursion set
theory for structure growth, and it contradicts a fundamental
assumption of the halo occupation distribution models often used to
study galaxy clustering, namely that the galaxy content of a halo of
given mass is statistically independent of its larger scale environment.
\end{abstract}

\begin{keywords}
methods: N-body simulations -- methods: numerical --dark matter --
galaxies: haloes -- galaxies: clusters: general
\end{keywords}
\title{The age dependence of halo clustering}

\section{Introduction}

Galaxy properties vary systematically with environment. Galaxies in
dense regions are more massive, more gas-poor, more bulge-dominated,
and have fewer young stars than those in low density regions. In
standard formation models, galaxies condense at the centres of
a hierarchically merging population of dark haloes (White \& Rees
1978).  Many recent models reproduce environmental effects by putting
early-type galaxies predominantly in massive haloes, late-type
galaxies in lower mass halos, while assuming the galaxy population in
a halo of {\it given} mass to be independent of where the halo lies
(e.g.  Kauffmann, Nusser \& Steinmetz 1997; Jing, Mo \& B\"orner,
1998; Benson et al 2000; Peacock \& Smith 2000; Wechsler et al 2001;
Berlind et al 2003; van den Bosch, Yang \& Mo 2003). This assumption
is justified by the standard excursion-set description of structure
formation (Bond et al 1991; Lacey \& Cole 1993; Mo \& White 1996)
where it follows from the Markov nature of the underlying random walks
(White 1996). It is also supported by the simulation results of Lemson
\& Kauffmann (1999) and Percival et al (2003) which detected no
dependence of halo clustering on properties such as concentration or
formation time. In contrast, the study of Sheth \& Tormen (2004) found
that ``haloes in dense regions form at slightly earlier times than
haloes of the same mass in less dense regions''.

In this Letter, we reevaluate the relation between environment and
formation history for dark haloes. In contrast with previous work, we
find the clustering at given mass to depend strongly on formation
time; low mass haloes that assemble early are much more strongly
clustered than those that assemble late. This dependence was missed in
most earlier numerical work because it is strongest at low masses.

The outline of our paper is as follows. In Section~2, we briefly
introduce the simulation used for our study. In Section~3, we compare
the mean halo bias measured from this simulation with the results of
others and with analytical models.  In Section~4, we present our
results for the dependence of spatial clustering on halo formation
time. Finally we give a short summary and discussion.

\section{The simulation}
The simulation used in this study is the so-called ``Millennium Simulation''
carried out by the Virgo Consortium (Springel et al. 2005). This simulation
adopted concordance values for the parameters of a flat $\Lambda$CDM
cosmological model, $\Omega_{\rm dm}=0.205$, $\Omega_{\rm b}=0.045$ for the
current densities in Cold Dark Matter and baryons, $h=0.73$ for the present
dimensionless value of the Hubble constant, $\sigma_8=0.9$ for the {\it rms}
linear mass fluctuation in a sphere of radius $8 h^{-1}$Mpc extrapolated to
$z=0$, and $n=1$ for the slope of the primordial fluctuation spectrum. The
simulation followed $2160^3$ dark matter particles from $z=127$ to the
present-day within a cubic region $500 h^{-1}$Mpc on a side. The individual
particle mass is thus $8.6\times 10^{8}h^{-1}{\rm M_\odot}$, and the
gravitational force had a Plummer-equivalent comoving softening of
$5h^{-1}$kpc. Initial conditions were set using the Boltzmann code
{\small CMBFAST} (Seljak \& Zaldarriaga 1996) to generate a realisation
of the desired power spectrum which was then imposed on a glass-like
uniform particle load (White 1996).

The {\small TREE-PM} N-body code {\small GADGET2} (Springel 2005) was used to
carry out the simulation and the full data were stored at $64$ times spaced
approximately equally in the logarithm of the expansion factor. This allowed
us to build trees which store detailed assembly histories for each of the
$5.7$ million dark matter haloes at $z=0$ that contain at least $64$
particles according to a {\small FOF} group finder with $b=0.2$ (Davis et al
1985). This is the set of haloes we analyse in the remainder of this paper.

\section{Mean halo bias as a function of mass and redshift}

Dark matter haloes are biased tracers of the underlying mass
density field, and models for the strength of this bias can be
constructed using the excursion set formalism of Bond et al (1991)
and Lacey \& Cole (1993).  Within this framework Mo \& White
(1996) and Sheth, Mo \& Tormen (2000) derived analytic expressions
for halo bias and tested them against N-body simulations of a
variety of cosmogonies (see also Cole \& Kaiser (1989)). The
linear density field can be represented by $\sigma(M, z)$, the
{\it rms} linear mass fluctuation (extrapolated to redshift $z$)
within a sphere which on average contains mass $M$. A
characteristic mass for clustering $M_*(z)$ can then be defined
through $\sigma(M_*, z) = \delta_c\approx 1.69$. Haloes more
massive than $M_*$ are predicted to be positively biased (i.e.
more strongly clustered) relative to the underlying mass, while
the opposite is true for less massive haloes. The tests in Mo \&
White (1996) showed the autocorrelation function for dark matter
haloes of mass $M$ to be approximately parallel to that of the
mass,
\begin{equation}
\xi_{hh}(r,M,z)=b^2(\nu,z)\xi_{mm}(r,z),
\end{equation}
where the ``bias factor'' $b(\nu,z)$ is given by
\begin{equation}
b(\nu,z)=1+(\nu^2-1)/\delta_c .
\end{equation}
Here $\delta_c$ is the critical linear overdensity at collapse and depends
slightly on cosmology; we use values from Eke et
al. (1996). $\nu=\delta_c/\sigma(M,z)$ is the dimensionless amplitude
of fluctuations that produce haloes of mass $M$ at redshift $z$. 

Recent high resolution $N$-body simulations have qualitatively
confirmed this model and the improved version of Sheth, Mo \& Tormen
(2000), but quantitative fits show some deviations (Jing 1998;
Governato et al 1998; Colberg et al 2000; Kravtsov \& Klypin 1999;
Seljak \& Warren 2004; Mandelbaum et al. 2005). Here we use the
unprecedented dynamic range and statistics of the Millennium
Simulation to study this bias relation further.

\begin{figure}
\centerline{\psfig{figure=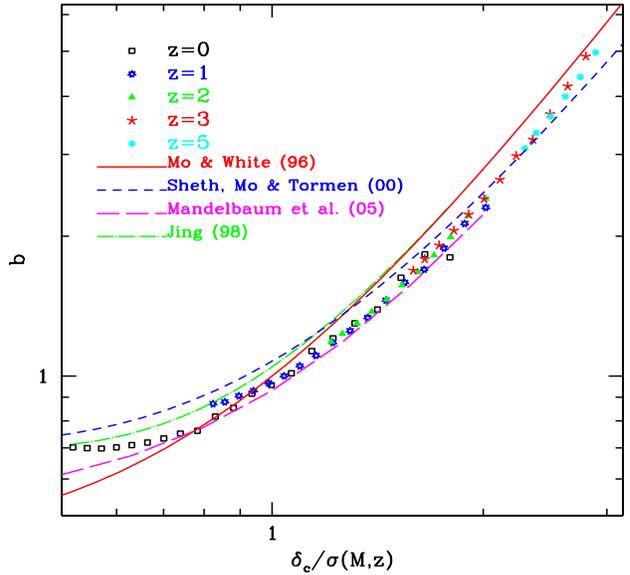,width=250pt,height=250pt}}
\caption{Halo bias as a function of peak height, $\nu=\delta_c/\sigma(M,z)$.
Individual symbols are the bias factors measured from the Millennium
Simulation; different symbols refer to different redshifts as indicated. The
red and blue lines are analytic predictions from Mo \& White (1996) and Sheth,
Mo \& Tormen (2000). The magenta and black lines are the fitting formulae
given by Jing (1998) and Mandelbaum et al. (2005). The latter are plotted only
over the parameter range covered directly by the numerical data.}
\label{fig:bias}
\end{figure}

For haloes of given mass and redshift, we derive a bias factor by comparing
their two point correlation function to that of the mass. More specifically,
we estimate $b^2(M, z)$ as the relative normalisation factor which minimizes
the mean square difference in $\log \xi$ for four equal width bins in $\log r$
spanning the separation range $6h^{-1}{\rm Mpc} < r < 25 h^{-1}{\rm Mpc}$. In
this range the measured correlations are all in the quasilinear regime.  Since
$b(M, z)$ is a steep function of $M$ for $M>M_*$, it is important to select
halo samples in relatively narrow mass ranges in order to
determine $b$ accurately at high mass. Here we choose mass bins of width $\Delta
M=0.3M$. The measured bias factors for the Millennium Simulation are shown by
the symbols in Fig.~\ref{fig:bias}. Note that we plot bias as a
function of peak height in order that the predicted relations for
different redshifts coincide. Different symbols denote the measured
bias factors for different redshifts. For comparison, we overplot the
analytic expressions which other authors have derived theoretically or
given as fits to their own numerical data.

\begin{figure}
\vspace{-0.5cm}
\centerline{\psfig{figure=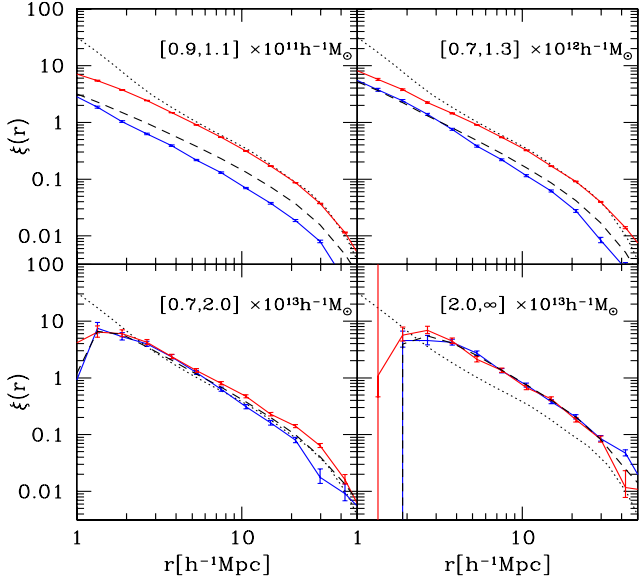,width=250pt,height=250pt}}
\caption{Two-point correlation functions for haloes in four mass
ranges.  Each panel gives results for haloes in the mass range
indicated in the label. The dotted black line, repeated in all
panels, is the correlation function of the underlying mass
distribution. Dashed black lines give the correlation functions
for the full sample of haloes in each mass range.  The red and
blue curves give correlation functions for the 20\% oldest and
20\% youngest of these haloes, respectively. Error bars are based
on Poisson uncertainties in the pair counts. Note that halo
exclusion effects are visible on small scales for the two most
massive samples.} \label{fig:typebias}
\end{figure}

The first noticeable feature is that when plotted in this way $b(\nu)$
depends at most weakly on redshift; at fixed $\nu$, the bias factors
for different redshift are almost identical provided that the critical
overdensity $\delta_c$ is calculated according to the recipe of Eke et
al (1996). Our results also agree well with previous work, lying
within the scatter of results from earlier large numerical
studies. Agreement is not perfect, however. As first noted by Jing
(1998), the analytic formula of Mo \& White (1996) overestimates the
bias factor at high mass and underestimates it at low mass. These
deficiencies are partially corrected by the ellipsoidal collapse
formula of Sheth, Mo \& Tormen (2000), but the scatter between the
numerical results of Jing (1998), Mandelbaum et al (2005) and this
paper is too large to allow any definitive conclusion. Possible
explanations for this scatter are differing numerical methods for
identifying haloes and for estimating $b$, differing numerical codes
to set initial conditions and integrate the evolution, different-sized
simulation volumes, differing assumed cosmologies, and analytic fits
which incompletely represent the numerical data. The scatter in our
figure is a measure of the remaining uncertainty in the mean halo
bias. It is much smaller than the systematic variation with halo
formation time which we turn to next.

\section{The dependence of clustering on formation time}
\begin{figure}
\resizebox{8cm}{!}{\includegraphics{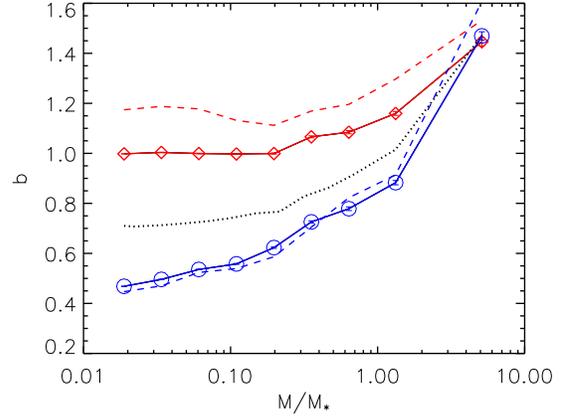}} \caption{Bias at
$z=0$ as a function of halo mass and formation time. Halo mass is
given in units of the characteristic mass ${\rm M_*} = 6.15\times
10^{12}h^{-1} {\rm M_\odot}$. The dotted black curve is the mean bias
for all haloes in the given mass bin. The solid red and blue
curves are for the 20\% oldest and 20\% youngest haloes,
respectively. The red and blue dashed curves refer to the 10\%
oldest and 10\% youngest haloes.}
\label{fig:timebias}
\end{figure}

\begin{figure*}
\hspace{8cm}\resizebox{16cm}{!}{\includegraphics{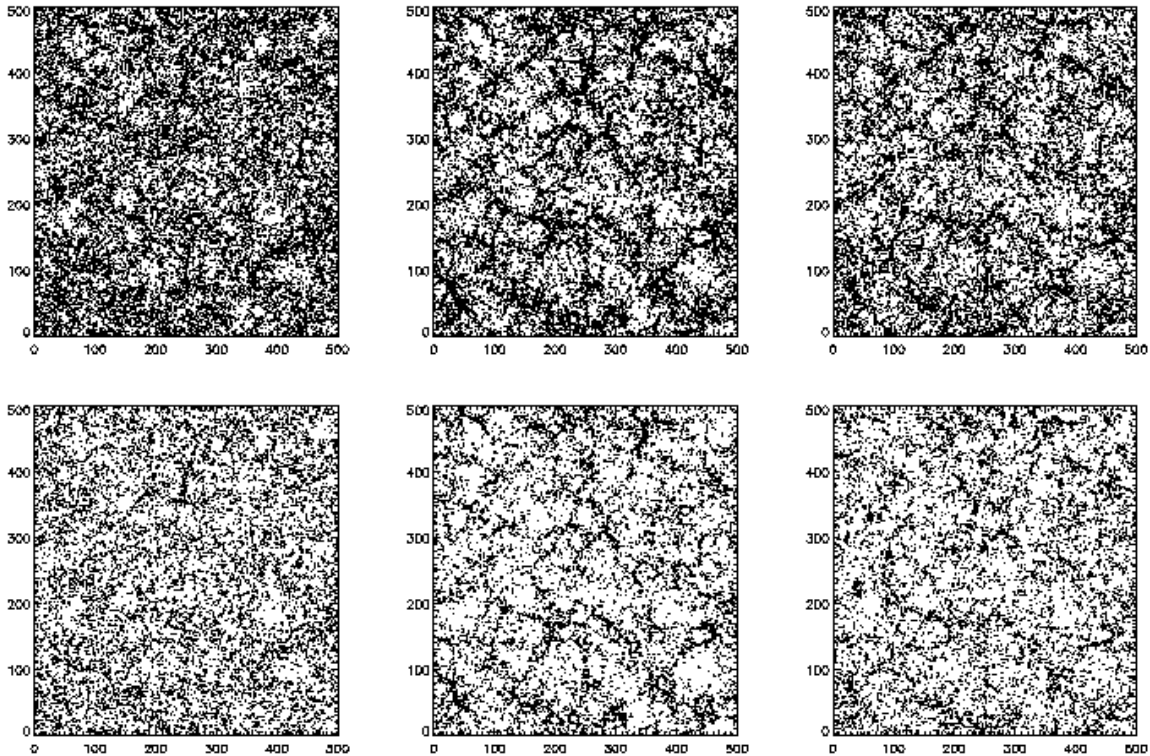}}
\caption{Images comparing the distribution of ``young'' haloes,
``old'' haloes and dark matter. The region plotted is a $30 h^{-1}
{\rm Mpc}$ slice through the Millennium Simulation. All haloes
plotted contain between 100 and 200 particles. The top row shows
the 20\% youngest (left) and 20\% oldest (middle) of these haloes,
together with an equal number of randomly selected dark matter
particles (right). The bottom row shows corresponding plots for
the 10\% tails of the halo formation time distribution.}
\label{fig:visual}
\end{figure*}

The formation time of a dark matter halo is conventionally defined as
the redshift when half of its final mass is first assembled into a
single object.  For each $z=0$ halo with more than 64 particles
(according to a standard {\small FOF} ($b=0.2$) group-finder) we follow
the stored merging tree to find the earliest time when the most
massive progenitor had more than half the final mass.  By linearly
interpolating its mass between this and the immediately preceding
output, we estimate the redshift when it had exactly half the final
mass. This we take as the halo formation time.

We examine the formation time dependence of the bias of $z=0$ haloes
as follows. For haloes in various mass bins we calculate two-point
correlation functions and bias factors both for the population as a
whole and for subsamples split according to formation time. We refer
to haloes with early formation times as ``old'' and to haloes with
recent formation times as ``young''. Fig.~\ref{fig:typebias} shows
results for four different mass ranges.  Each panel compares the
autocorrelation function for all the haloes in the given mass range
both with that of the underlying dark matter and with autocorrelation
functions for subsamples made up of the 20\% oldest and 20\% youngest
haloes. For haloes of $10^{11}$ and $10^{12}h^{-1}{\rm M_\odot}$ the
dependence of clustering on formation time is strong and increases
systematically with increasing length-scale. The effect is detectable
but weak at $10^{13}h^{-1}{\rm M_\odot}$. It is undetectable for haloes more
massive than $2\times 10^{13}h^{-1}{\rm M_\odot}$.

In Fig. ~\ref{fig:timebias} we show more directly how the formation time
dependence of clustering varies with halo mass. We divide the $z=0$ haloes
into a series of mass bins of width $\Delta\log M=0.3$. In each bin we
estimate bias factors as described above for the halo population as a whole
and for the 10 and 20\% tails of oldest and youngest haloes.  The plot shows
clearly that the relative bias of old versus young haloes increases smoothly
with decreasing halo mass. The bias of the 10\% youngest haloes is only
slightly stronger than that of the 20\% youngest haloes, but the 10\% oldest
haloes are significantly more clustered than the 20\% oldest haloes. The
effects become very large for the lowest masses that we resolve. At
$10^{11}h^{-1}{\rm M_\odot}$ the large-scale autocorrelation amplitude for the 10\%
oldest haloes is more than 5 times that for the 10\% youngest haloes.

In Fig.~\ref{fig:visual}, we provide some images to give a visual
impression of the relative distributions of ``young'' and ``old''
haloes.  Here we show haloes with {\small FOF} particle number in the
range $[100, 200]$ in a slice through the Millennium Simulation $30
h^{-1} {\rm Mpc}$ thick. The top row shows the positions of the 20\%
youngest of these haloes (left), of the 20\% oldest (middle), and of
an equal number of dark matter particles selected at random within the
slice (right). The bottom row shows corresponding plots for the 10\%
tails.  It is striking that although the haloes (by definition) avoid
massive clumps in the dark matter distribution, the old haloes follow
the large-scale cosmic web quite closely, while the distribution of
young haloes looks almost uniform.

\begin{figure}
\resizebox{8cm}{!}{\includegraphics{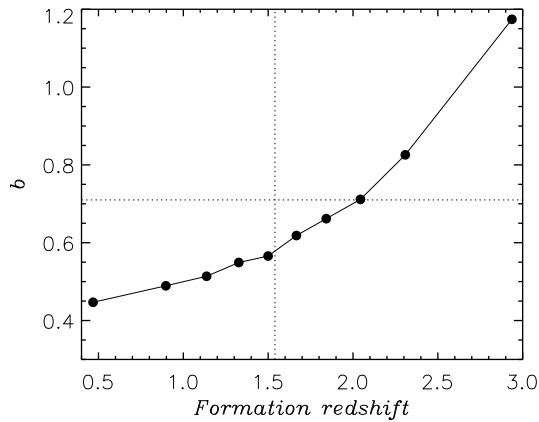}} \caption{Bias as a
function of halo formation time.  We divide haloes with particle
number in the range $[100,200]$ into ten equal-sized subsamples as
a function of their formation time. For each subsample we compute
a mean formation redshift and a bias factor. The figure plots
these two quantities against each other.  Vertical and horizontal
dotted lines show the mean formation redshift and the mean bias
for the sample as a whole.} \label{fig:biasinbin}
\end{figure}

Fig. ~\ref{fig:biasinbin} explores the formation time dependence
of the clustering bias in more detail. We take the sample of all
haloes with particle number in the range $[100, 200]$ and we split
it into ten equal-sized subsamples by formation time. We then
compute bias factors and mean formation redshifts for each of
these subsamples and plot one against the other. While the
variation of bias with formation time is smooth, the strongest
effects clearly occur for the ``oldest'' haloes. Notice also that
the variation in mean formation redshift is large, ranging from
$z=0.47$ for the youngest 10\% to $z=2.94$ for the oldest 10\%.
The mean formation redshift for the population as a whole is
$z=1.54$.

\section{Discussion and conclusions}
In this Letter, we have used the very large Millennium Simulation
(Springel et al 2005) to study how the clustering of dark haloes
depends on mass and formation time. Our results for the mean
dependence of bias on mass agree well with those of other workers, but
for low mass haloes, $M\leq M_*$, we find a strong and unexpected
dependence on formation time.  Haloes of given mass that assembled at
high redshift are substantially more strongly clustered than haloes of
the {\it same} mass that assembled recently. This difference persists
to large scale, and indeed is larger on scales of order $20h^{-1}$Mpc
than on scales of order $2h^{-1}$Mpc. Although there is unavoidably
some arbitrariness in our definitions of halo mass and formation time,
experiments with alternative definitions of each quantity produce bias
variations which are typically of order 10\%. Our basic result thus
appears robust.

This result is unexpected because the sharp $k$-space filter used in
most formulations of the excursion set model for structure formation
(Bond et al 1991; Lacey \& Cole 1993) causes the random walks as a
function of smoothing scale on which the model is based to have
Markovian character. The formation history of a halo is encoded in the
random walk at higher mass resolution than that which defines the halo
itself, and is thus statistically independent of the environment,
which is encoded in the random walk at lower resolution (White
1996). This independence of history and environment appeared confirmed
by the simulation analyses of Lemson \& Kauffmann (1999) and Percival
et al.(2003).  These concentrated on relatively massive objects for
which the effects we find are small, probably undetectable with
volumes of the size analysed. A formation time dependence was seen,
however, in the mark correlation analysis of Sheth \& Tormen (2004),
although these authors characterised the effect as ``slight''.

Independence of history and environment for haloes of given mass
is assumed in many theoretical models for galaxy formation. For
example, Kauffmann, Nusser \& Steinmetz (1997), Benson et al (2000)
and Wechsler et al (2001) populate haloes in dark matter simulations
by using semi-analytic models applied to a Monte Carlo realisation of
each halo's history depending only on its mass. Halo occupation
distribution models such as those of Jing, Mo \& B\"orner (1998),
Peacock \& Smith (2000), Berlind et al (2003) or van den Bosch, Yang
\& Mo (2003) ignore formation histories altogether and assume {\it a
priori} that the galaxy population of a halo depends on mass alone,
independent (at {\it fixed} mass) of the larger scale environment.

In practice, haloes similar in mass to that of the Milky Way contain a
substantial fraction of the galaxies in typical observational
surveys. Since it is plausible that galaxxy properties should depend
significantly on the assembly history of their haloes, our results
suggest that models which ignore the age dependence of halo clustering
will incorrectly predict the large-scale distribution of galaxies. The
extent of the problem may depend on the specific galaxy formation
model considered.  We will estimate its size for a typical
``successful'' model in a future paper.

Finally we note that galaxy formation models which explicitly
follow the assembly history of each halo should be immune to this
problem. Examples of such models which graft a semi-analytic
treatment of baryon physics onto a N-body simulation can be found
in Kauffmann et al (1999), Springel et al (2001, 2005), Helly et
al (2003), Hatton et al (2003) and Kang et al (2005).
Cosmological simulations which follow the baryonic physics
directly should also be safe, although it is not yet
computationally feasible to simulate galaxy formation reliably
throughout a big enough volume for the effects discussed here to
dominate the errors.

\section*{Acknowledgements}
We are grateful to Adrian Jenkins and Carlos Frenk for detailed
comments on our paper. GL also thanks Yipeng Jing and Darren Croton
for useful discussions.  The simulation used in this paper was carried
out as part of the programme of the Virgo Consortium on the Regatta
supercomputer of the Computing Centre of the Max-Planck-Society in
Garching.
\label{lastpage}

\end{document}